\documentclass[12pt]{article}
\pdfoutput=1
\usepackage{jhep-mod}
\usepackage{bm}
\usepackage{amssymb}
\usepackage{pifont}
\usepackage{slashed} 
\usepackage{slantsc} 
\usepackage[applemac]{inputenc}
\title{Thermality of the Hawking flux}
\author{Matt Visser}
\affiliation{School of Mathematics, Statistics, and Operations Research, \\
Victoria University of Wellington, PO Box 600, Wellington 6140, New Zealand}
\emailAdd{matt.visser@msor.vuw.ac.nz}

\abstract{
Is the Hawking flux ``thermal''?  
Unfortunately, the answer to this seemingly innocent question depends on a number of often unstated, but quite crucial, technical assumptions built into modern (mis-)interpretations of the word ``thermal''.  
The original 1850's notions of thermality --- based on classical thermodynamic reasoning applied to idealized ``black bodies''  or ``lamp black surfaces'' --- when supplemented by specific basic quantum ideas from the early 1900's, immediately led to the notion of the black-body spectrum, (the Planck-shaped spectrum), but \emph{without} any specific assumptions or conclusions regarding correlations between the quanta. 
Many (not all) modern authors (often implicitly and unintentionally) add an extra,  quite unnecessary, assumption that there are no correlations in the black-body radiation; but such usage is profoundly ahistorical and dangerously misleading. 
Specifically, the Hawking flux from an evaporating black hole, (just like the radiation flux from a leaky furnace or a burning lump of coal), is only \emph{approximately} Planck-shaped over an explicitly bounded range of frequencies.  
Standard physics (phase space and adiabaticity effects) explicitly bound the frequency range over which the Hawking flux is \emph{approximately} Planck-shaped  from both above and below --- the Hawking flux is certainly not \emph{exactly} Planckian, and there is no compelling physics reason to assume the Hawking photons are uncorrelated.

\bigskip
\noindent
27 September 2014; 9 February 2015; 7 May 2015; \LaTeX-ed \today
}
\begin{document}
\maketitle
\def\R{{\mathbb{R}}}
\def\d{{\mathrm{d}}}
\def\sech{{\mathrm{sech}}}
\def\O{{\mathcal{O}}}
\def\Dirac{{\slashed D }}
\providecommand{\ceil}[1]{\left \lceil #1 \right \rceil }
\clearpage
\section{Introduction}
\label{S:Introduction}

Stephen Hawking predicted, (now some 40 years ago), that semi-classical black holes will emit quantum radiation with a temperature proportional to their surface gravity, and will slowly evaporate due to subtle quantum effects~\cite{hawking74a, hawking74b, hawking76}.  
Direct experimental tests of this phenomenon have so far been impractical, and the best laboratory data comes from analogue systems such as surface waves in a water tank~\cite{Weinfurtner:2010, Weinfurtner:2013}, and more recently, phonons in a BEC~\cite{Steinhauer:2014}.
Faced with this extreme paucity of both experimental and observational data, the community has focussed almost entirely on \emph{gedanken-experiments}, with well over 5000 theory papers generated to date.

\enlargethispage{20pt}

While there is almost universal agreement that the predicted Hawking flux will actually occur for a general relativistic black hole, and almost universal agreement that the associated back-reaction will slowly reduce the black hole's mass, there is relatively little agreement (in fact, considerable disagreement) regarding the endpoint of the Hawking evaporation process. 
In particular, the question of the thermality of the Hawking flux, (and the \emph{precise sense} in which it is thermal), is a crucial and  important ingredient in the so-called  ``information puzzle'', and its more recent ``firewall'' variant~\cite{firewall, firewall2, firewall:2009, no-firewall, no-firewall2, no-firewall3, no-firewall4, no-firewall5, no-firewall6, no-firewall7, no-firewall8, no-firewall9}; this has important implications regarding the quantum unitarity of the Hawking evaporation process~\cite{Page:1979, Hawking:1982, Banks:1983, Preskill:1992, Hawking:1995}. 
A key point is this: There is a crucial difference between the ``qualitative" and ``quantitative" information loss problems.
\begin{itemize}
\item 
The ``qualitative" problem is this: If a spacelike singularity forms (in the strict mathematical sense), then there will be a (strict mathematical) event horizon,  and unavoidably \emph{some} loss of unitarity associated with any matter that might cross the \emph{event} horizon.\footnote{
In his 1976 article~\cite{hawking76} on the breakdown of predicability, Hawking phrased the discussion in terms of ``hidden surfaces'', reserving the phrase ``event horizons'' specifically for black holes. By doing things in this way his discussion applied also to the branching-off of baby universes, or indeed any sort of nontrivial temporal topology associated with what is now typically called a ``causal horizon''. 
For the purposes of this current article I will focus on the event horizons (possibly) associated with physical black holes, and the  apparent/trapping horizons definitely associated with physical black holes~\cite{observability}. 
Hawking's 1976 argument will apply whenever there is a causally inaccessible region in which one can hide correlations. 
Certainly that argument applies to the event horizons (possibly) associated with physical black holes.}
\item 
The ``quantitative" problem is this: \emph{How much} information is lost behind the event horizon, (if it forms), and how much comes out in the Hawking radiation? 
\end{itemize}
Many authors have argued that \emph{no} information comes out in the Hawking radiation, but that is rather begging the question. Other authors only address part of the puzzle. This subtlety has often been lost in the sometimes heated exchange of comments and criticisms.

\clearpage
\noindent
There is another way of saying this: 
\begin{itemize}
\item 
Hawking radiation is associated with the apparent/trapping horizon, and couldn't care less about the event horizon (if present).
\item
Unitarity violation (if present) is associated with the event horizon (if present), and couldn't care less about the apparent/trapping horizon. 
\end{itemize}
Only if you \emph{assume} that the event horizon actually forms, \emph{and that it closely tracks the apparent/trapping horizon},
is there ever any significant information loss. 
To expand on these issues, let us first consider the spectral shape, and then the correlation structure, of the Hawking flux.  

\section{Spectrum of the Hawking flux}
Hawking's 1974 derivation~\cite{hawking74a,hawking74b}, the many and quite varied subsequent re-derivations thereof, and the modern \emph{adiabatic} variants of Hawking's original calculation~\cite{minimal1, minimal2}, all agree that the shape of the spectrum is \emph{approximately} a Planckian  black-body spectrum, \emph{but with certain key modifications and limitations}. 
Specifically, the Planckian shape of the Hawking spectrum will, \emph{at an absolute minimum}, be modified by \emph{at least} three distinct physical effects: 
\begin{enumerate}
\item 
greybody factors;
\item 
adiabaticity constraints; 
\item
available phase space. 
\end{enumerate}
Let us consider these three effects in turn.

\bigskip
 {\bf--- Greybody factors:} 
 These well-known effects arise when the Hawking flux is back-scattered by the non-trivial gravitational field between the quasi-local horizon (the apparent/trapping horizon) and spatial infinity; the existence of these greybody effects is entirely standard and well-known, though quantitative estimates are sometimes tricky. 
 Early work dates back to the mid-to-late 1970's~\cite{Page:1976a, Page:1976b, Bekenstein:1977}, and interest in these quantities is active and ongoing, see for instance~\cite{Escobedo:2008, Schwarzschild, Static-Spherically-Symmetric}.  
 A key feature is to note that the greybody factors will always \emph{suppress} the Hawking flux.
 
\bigskip 
{\bf--- Adiabaticity constraints:} 
These lesser-known effects arise from including back-reaction, and depend on the fact that the spacetime geometry must be slowly evolving (on the time-scale set by the frequency of the Hawking photon) in order for Hawking's calculation, or any of its more modern variants, to apply.  
Implications of the adiabaticity condition were carefully analyzed and discussed in both references~\cite{minimal1} and~\cite{minimal2}.
Specifically, to obtain an approximately Planckian spectrum the surface gravity must be ``slowly evolving'' in the sense that:
\begin{equation}
|\dot \kappa| \ll \kappa^2.
\end{equation}
This is the constraint that a photon at the peak of the Hawking spectrum should \emph{not} see any significant change in the surface gravity during one oscillation period. 

More generally and quantitatively, let us now consider the conditions for the validity of Hawking's ``exponential approximation'' for the relative $e$-folding of the affine null parameters between past and future null infinity (scri$^-$ and scri$^+$)~\cite{hawking74a, hawking74b}. 
Whenever the surface gravity is time-varying, $\dot \kappa\neq 0$, then the exponential approximation is at best valid over a bounded time interval of width~\cite{minimal1, minimal2}:
\begin{equation}
\Delta t \ll {1\over\sqrt{|\dot \kappa|}}.
\end{equation}
See the discussion surrounding equation (10) of reference~\cite{minimal1} or equation (3.22) of reference~\cite{minimal2} for extensive technical details. 
But then Hawking's argument, (or its more modern variants), can be applied only to wave packets which can be localized within this time interval. 
This implies the wave-packet must be built out of modes of frequency  \emph{at least}  $\omega_\mathrm{min}= \sqrt{|\dot \kappa|}$.
In particular, a suitable extension of Hawking's argument leading to a Planckian spectrum only works for 
the limited range of frequencies 
\begin{equation}
\omega  \gtrsim  \omega_\mathrm{min}= \sqrt{|\dot \kappa|}.
\end{equation}
Thus this adiabaticity argument provides an \emph{infrared} frequency cutoff on the Hawking flux. 
If Hawking's argument (suitably extended) is to apply not just in the exponential Boltzmann tail, but also to include the peak of the Planck spectrum, then one recovers the quantitative condition $|\dot \kappa| \ll \kappa^2$.

\bigskip
{\bf--- Phase space effects:} 
Though typically ignored, there is also a phase space \emph{ultraviolet} frequency cutoff in the Hawking flux.  
At its crudest, the emitted photon energy can never exceed the available mass energy: $\omega< m$.  
A slightly safer statement, for charged or rotating black holes, is $\omega < m - m_\mathrm{extremal}$. 

More carefully, applying (standard flat-space) Lorentz kinematics in the far distant asymptotically flat region, it is easy to see that for a black hole of mass $m_i$ emitting, (in its rest frame), a photon of energy $\omega$, and thereby reducing its mass to $m_f$, one has:
\begin{equation}
\omega =  {m_i^2-m_f^2\over2m_i}.
\end{equation}
Applied to the photons in the Hawking flux this yields the purely kinematic bound:
\begin{equation}
\omega \leq  {m^2-m_\mathrm{extremal}^2\over2m}.
\end{equation}
Thus this argument provides an \emph{ultraviolet} frequency cutoff on the Hawking flux.\footnote{
My own early views on the importance of the phase-space cutoff can be found in reference~\cite{Visser:1992}.
 Although I am no longer in favour of the particular way that I discretized black hole entropy in that article, the comments regarding the importance of the phase-space cutoff and the final ``particle cascade'' leading to complete evaporation of Planck-scale black holes still hold.}$^{,}$%
\footnote{
More recently the Parikh--Wilczek approach to Hawking radiation viewed as quantum tunnelling also explicitly (but somewhat indirectly) includes at least some phase-space effects and also adds nonlinear frequency-dependent terms in the action~\cite{Parikh:1999}. 
Parikh and Wilczek consider the emission of spherically symmetric thin shells, so 3-momentum conservation is trivial, and the phase space cutoff simplifies to $\omega \leq m- m_\mathrm{extremal}$. 
For the Schwarzschild black hole  Parikh and Wilczek find: $\mathrm{Im}(\mathrm{Action}) =  4\pi \omega m (1-{\omega\over2m})$, and relate this to $\Delta (\mathrm{Entropy}) =  8\pi \omega m (1-{\omega\over2m})$, subject to $\omega\leq m$. 
Some authors prefer to interpret this as a frequency-dependent temperature, $T_\mathrm{effective}(\omega) =  T_\mathrm{Hawking} \times \left( 1 - {\omega \over 2m }\right)^{-1}$. 
The situation for Reissner--Nordstr\"om black holes is considerably more subtle. 
There 
\[
\Delta (\mathrm{Entropy})  = 2\pi\left\{\omega (2m-{\omega})-(m-\omega)\sqrt{(m-\omega)^2-q^2} + m\sqrt{m^2-q^2}\right\}, 
\quad
\hbox{subject to }\omega \leq m - |q|.
\]
If desired, an effective temperature can be defined by $T_\mathrm{effective}(\omega) = \omega/\Delta(\mathrm{Entropy})$, with a low-frequency expansion  $T_\mathrm{effective}(\omega) =  T_\mathrm{Hawking} + {\mathcal O}(\omega)$. 
Thus the Parikh--Wilczek approach provides both an explicit phase-space cutoff, and a modified emission amplitude.
}

\bigskip
{\bf--- Combined effects:} 
Combining these adiabaticity and phase space constraints, and noting the existence of grey-body effects,  we can make the quantitative statement that the Hawking flux can (at best) be \emph{approximately} Planckian only over the limited frequency range:
\begin{equation}
\omega \in \left(  \sqrt{|\dot\kappa|}, \;  {m^2-m_\mathrm{extremal}^2\over2m} \right).
\end{equation}
Even within this range, where the (suitably extended modern variants of the) Hawking calculation can be trusted,  greybody factors (barrier transmission probabilities) will to some extent \emph{suppress} the Hawking flux below that of an ideal  Planck spectrum.

\section{Schwarzschild black holes}

Let us now see what this quantitatively implies for Schwarzschild black holes: 
For the specific case of the Schwarzschild black hole $m_\mathrm{extremal}=0$, so the phase space cutoff is simply $\omega<{m/2}$. Indeed, for Schwarzschild black holes the phase space cutoff never intersects the peak ($\omega\sim\kappa$) of the approximately Planck-shaped spectrum while one remains within the semi-classical regime.

Because the phase space cutoff is so high, (compared to the location of the Hawking peak at $\omega\sim\kappa$), it is perfectly acceptable, (at least as a zeroth-order approximation), to approximate the Hawking flux by a complete Planck spectrum, integrate over all of phase space, and so get the Stefan--Boltzmann law,  ($\dot m = -\sigma \;T^4 \;A_\mathrm{horizon}$). 
But I emphasise that the applicability of the Stefan--Boltzmann law is intrinsically  an \emph{approximate} result; in view of the physical arguments presented above it \emph{cannot} be exact.  

Then, introducing  Planck quantities for simplicity, we have $\kappa \sim {m_P^2/m}$.  
Thereby we deduce:
\begin{equation}
\dot\kappa = - {\dot m m_P^2\over m^2} = \left\{ \left(m_P^2\over m\right)^4 \times 4\pi (2m)^2 \right\} {m_P^2\over m^2}
\sim {m_P^6 \over m^4}. 
\end{equation}
Consequently the Hawking flux from a Schwarzschild black hole is approximately Planckian (up to greybody factors) over the rather broad frequency interval
\begin{equation}
\omega \in \left( {m_P^3 \over m^2} , \;  {m\over2} \right).
\end{equation}
This interval is certainly non-empty for macroscopic black hole masses, and even for mesoscopic black hole masses all the way down to the Planck scale.
Furthermore
\begin{equation}
{|\dot\kappa|\over\kappa^2} \sim {m_P^6 / m^4\over m_P^4/m^2} = {m_P^2\over m^2}.
\end{equation}
So, as claimed, the peak of the Planck blackbody spectrum is indeed contained in the approximately Planckian interval all the way down to the Planck scale, (where one should stop believing semiclassical physics anyway).

Thus we see that for a Schwarzschild black hole these three bounds on the Planckian nature of the Hawking flux are (numerically) not particularly stringent. But they do however provide important \emph{qualitative information} --- at the very least they serve as a suitable antidote to the often made, (and often repeated, but utterly incorrect), assertion that the Hawking flux is \emph{exactly} Planckian.

\section{Connecting future and past null infinities} 
Let $U$ be an affine coordinate on past null infinity, while $u$ is taken to be an affine coordinate on future null infinity. 
Much of the physics of the Hawking effect is encoded in the $e$-folding relation connecting past and future null infinities~\cite{hawking74a, hawking74b}
\begin{equation}
U = U_H - A \; \exp(-\kappa_H u).
\end{equation}
Once one includes the effects of a time-dependent evolving black hole one should instead write~\cite{minimal1, minimal2}
\begin{equation}
U(u) = U_0 + \int_{u_0}^u \exp\left(-\int_{u_0}^{\bar u} \kappa(\tilde u) \; \d \tilde u\right) \; \d \bar u.
\end{equation}
Here $u_0$ is merely some convenient starting point, often taken to be the onset of black hole formation. 
If we approximate $\kappa(u)\approx\kappa_0$ as a constant then
\begin{equation}
U(u) \approx U_0 + {1\over\kappa_0} -  {\exp(-\kappa_0[u-u_0])\over\kappa_0},
\end{equation}
which is equivalent to the naive result used in the original 1974 calculations~\cite{hawking74a, hawking74b}.

\bigskip
Now let us make this more explicit and quantitative: 
When including the effects of back-reaction, for an evolving Schwarzschild black hole of initial mass $m_0$ we have $\kappa_0 \sim m_P^2/m_0$, and from the previous section $\dot\kappa \sim m_P^{-2} \kappa^4$. We can write this more carefully as the exact scaling relations $\kappa = \kappa_0 (m_0/m)$ and $\dot\kappa = \dot\kappa_0 (\kappa/\kappa_0)^4$, or even $\dot\kappa(u) = B \kappa(u)^4$, where approximately $B\sim m_P^{-2}$.  
Thereby 
\begin{equation}
\kappa(u) = {\kappa_0\over \sqrt[3]{1-3 B \kappa_0^3  [u-u_0]}}.
\end{equation}
This approximation will remain valid until the surface gravity rises to the Planck scale, which will happen when
\begin{equation}
{m_P^3\over m^3} \sim 1-3 [u-u_0]{m_P^4\over m^3}.
\end{equation}
That is when
\begin{equation}
u - u_0 \sim {T_P\over 3} \left( {m^3\over m_P^3} - 1\right) \sim {T_P\over 3}  {m^3\over m_P^3}.
\end{equation}
During that entire interval, from $u=u_0$ to $u\sim u_0+ {1\over 3} T_P (m^3/ m_P^3)$, the surface gravity (while not constant) is still slowly varying, in the sense of satisfying the adiabaticity constraint. 
In this interval $\kappa(u)$ can be  integrated to explicitly yield
\begin{equation}
\int_{u_0}^u \kappa(u) \; \d u = {1\over 2 B \kappa_0^2} \left[ 1 -  \sqrt[3]{1-3 B \kappa_0^3  [u-u_0]}\right].
\end{equation}
A second integration now gives
\begin{eqnarray}
U(u) = U_0 &+& {1\over\kappa_0}  -  \sqrt{2B} \; D\left( 1\over\sqrt{2B} \kappa_0\right)
 \nonumber\\
&-& 
 {1\over\kappa(u)}\exp\left({1\over2B} \left[ {1\over\kappa(u)^2} - {1\over\kappa_0^2} \right]\right) 
\nonumber \\
 &+& 
\sqrt{2B}  \exp\left({1\over2B} \left[ {1\over\kappa(u)^2} - {1\over\kappa_0^2} \right]\right) \; D\left(1\over\sqrt{2B} \kappa(u) \right) .
\end{eqnarray}
Here $D(x)$ is the \emph{Dawson function}
\begin{equation}
D(x) = e^{-x^2} \int_0^x e^{t^2} \d t. 
\end{equation}
If desired the Dawson function can (up to rescaling) be related to the error function for imaginary argument, but in this real form is more suited to numeric manipulations.  
In particular the Dawson function is bounded by $D(x) < 0.54105$, which means that the terms involving the Dawson function never shift $U(u)$ by more than one Planck time, and so can quietly be neglected until one reaches the Planck regime.  

So for all practical purposes
\begin{eqnarray}
U(u) = U_0 + {1\over\kappa_0}  -
 {1\over\kappa(u)}\exp\left({1\over2B} \left[ {1\over\kappa(u)^2} - {1\over\kappa_0^2} \right]\right)  + \mathcal{O}(T_P).
\end{eqnarray}
Note that while the relationship between future and past null infinity is now quite considerably more complicated than the simple $e$-folding of references~\cite{hawking74a,hawking74b}, we see that it is nevertheless quite explicit. 
Furthermore, at any particular time $u_*$ one can always locally approximate the exact $U(u)$ with an $e$-folding expression of the form~\cite{minimal1,minimal2}
\begin{equation}
U(u\approx u_*) \approx U_* + {1\over\kappa(u_*)} -  {\exp(-\kappa(u_*) \; [u-u_*])\over\kappa(u_*)}.
\end{equation}
This formalism now gives one a slowly evolving Hawking temperature, at least until the mass of the black hole drops sufficiently low so that one enters the Planck regime.

\section{Effective temperature} 
The net effect of these greybody factors, adiabaticity constraints, and phase space constraints is to modify the spectrum of the Hawking flux:
\begin{equation}
n(k) = {f(\omega)\over\exp(\hbar\omega/k_B T_H) -1}; \qquad f(\omega) \in (0,1).
\end{equation}
Here $f(\omega)$ is some dimensionless suppression factor now encoding all three effects. 
This allows one (in the quite usual manner) to define an effective temperature in terms of the total energy flux:
\begin{equation}
\sigma \; T_\mathrm{effective}^4 = \int n(k) \;\hbar\omega\; {\d^3 k\over (2\pi)^3}. 
\end{equation}
Setting $z=\hbar\omega/k_B T_H$, and $f(\omega) \to f(z)$, one sees
\begin{equation}
 T_\mathrm{effective}^4  = {\displaystyle\int {f(z) z^3\over e^z-1}\; \d z\over\displaystyle \int { z^3 \over e^z-1}\; \d z} \; \;T_H^4 
 \;\;\leq\;\;T_H^4. 
\end{equation}
That is, $T_\mathrm{effective} \leq T_H$, so the effective temperature (bolometrically defined) of the outgoing Hawking flux has been suppressed below the naive Hawking temperature.

\bigskip
This then changes (both qualitatively and quantitatively) the entropy budget in the Hawking evaporation process. 
As the black hole evaporates, its Bekenstein entropy~\cite{bekenstein} decreases as 
\begin{equation}
\d S_B = -{|\d M|\over T_H},
\end{equation}
whereas the entropy change of the outgoing radiation can best be estimated as 
\begin{equation}
\d S_H = +{|\d M|\over T_\mathrm{effective}}.
\end{equation}
(The outgoing radiation, since it is not exactly Planckian, should really be analyzed using non-equilibrium thermodynamics; but use of the effective temperature is a well-known stand-in for such effects.)
Overall one has
\begin{equation}
\d S_\mathrm{total} =  |\d M| \; \left({1\over T_\mathrm{effective}} - {1\over T_H}\right) \geq 0.
\end{equation}
So the Hawking evaporation process actually \emph{increases} the total entropy of the universe. 
(Note this is intrinsically a coarse-graining entropy associated with throwing away detailed information regarding the Hawking flux; this argument has nothing to say one way or another regarding the unitarity of the underlying physical process.) 
This is perhaps somewhat unexpected from the standard point of view, but is utterly unavoidable as soon as one takes proper cognisance of greybody, adiabaticity, and phase space effects.

\section{Wick rotation and the Hawking flux} 
Ultimately the origin of the often-made but mistaken assertion that  the Hawking flux is \emph{exactly} Planckian seems to trace back to an over-enthusiastic and uncritical adoption of Wick rotation (Euclidean quantum gravity) techniques~\cite{Hawking:Euclidean}.  
Certainly the Wick rotation of a static black hole ($t\to i t$, in the manifestly static coordinate system where $g_{ti}=0$), combined with the condition that there be no conical singularity at the Euclideanized version of the Lorentzian-signature Killing horizon, picks out the surface gravity as being physically important,  being related (via periodicity in imaginary time) to a notion of temperature --- but this is by construction an \emph{intrinsically equilibrium} argument for a black hole in \emph{exact thermal equilibrium with a heat bath} at the Hawking temperature~\cite{Hawking:Euclidean}. 
\begin{itemize}
\item 
By construction the heat bath has an exactly Planckian spectrum, simply because it is \emph{assumed} to be in exact thermal equilibrium; the greybody factors quietly drop out. 
\item
By construction the situation is static; there simply are no adiabaticity conditions since $\dot\kappa\equiv0$ exactly. 
\item
By construction there are no phase-space constraints; since (typically) one is completely ignoring back-reaction. 
\end{itemize}
But this Wick-rotated Euclideanized system tells you relatively little regarding the non-equilibrium emission of the Hawking flux into vacuum; the Unruh quantum vacuum state is radically different from the Hartle--Hawking quantum vacuum state.  
(Wick rotation automatically puts one into the Hartle--Hawking quantum vacuum state, not the physically relevant Unruh vacuum state.) 
While the Wick rotation trick provides a ``quick and dirty'' way of relating surface gravity to Hawking temperature~\cite{temperature}, it misses much of the essential physics of the evaporation process.  
Once one considers a real black hole evaporating into vacuum, the Hawking flux is no longer exactly Planckian --- the shape of the spectrum must at the very least be modified by the three physical effects considered above.

\section{Correlations in the Hawking flux} 
Are the Hawking quanta in any way correlated with each other? 
This quite deceptively innocent question can easily initiate a firestorm of quite inconclusive debate. 

The original 1850's notions of thermality, based as they were on entirely classical thermodynamic reasoning applied to black bodies, (such as, for instance, the traditional ``leaky cavity'' or ``lamp-black'' surfaces),  made no intrinsic assumptions regarding the possibility of correlations in the outgoing radiation.  
But modern abuse of the word ``thermal'' often implicitly makes assumptions about a lack of correlations.
It is essential to realise that the physical distinction between ``Planckian'' (Planck-shaped spectrum) and ``thermal'' is both important and subtle.
No-one seriously doubts that burning a lump of coal in a leaky furnace results in an approximately Planckian spectrum, (an approximately ``black body'' spectrum), nor that this process implies correlations in the outgoing radiation --- which  then cannot be exactly ``thermal'' in the technical sense that this word has come to be used (or rather abused) in the modern literature. 

Now Hawking's original 1974 calculation~\cite{hawking74a,hawking74b}, (and its modern adiabatic variants, see for instance references~\cite{minimal1} and~\cite{minimal2}), certainly demonstrate that a collapsing ball of matter will excite the quantum vacuum state, and that the outgoing radiation is approximately blackbody, that is, has an approximately  Planck-shaped spectrum,  (at least up to greybody, adiabaticity, and phase space effects, as discussed above). 
But the considerably stronger statement that the outgoing Hawking quanta are completely uncorrelated, (the modern misuse of the word ``thermal''), depends on a separate and very much stronger implicit assumption: 
That in a semi-classical astrophysical black hole an \emph{event} horizon forms to \emph{permanently} hide any possible correlations, in such a way that they never again become visible to the external universe --- but the possibility of doing this depends on delicate issues of global  geometry --- including what will happen in the infinite future~\cite{observability}. 
In contrast \emph{apparent} horizons or \emph{trapping} horizons, while they may temporarily hide correlations, do not necessarily do so permanently.
Without an \emph{event} horizon, whose very existence is delicately predicated on \emph{assumptions} being made about the infinite future, a black hole defined in terms of \emph{apparent} or \emph{trapping} horizons will behave much more like a furnace; a leaky furnace with a small hole in it, the original 1850's classical thermodynamic definition of a ``black body''.

\section{Analogue Hawking flux} 
To really drive home the point that the existence of possible correlations in the Hawking flux is logically independent from  the existence of the Hawking flux itself simply consider an acoustic black hole (dumb hole)~\cite{Unruh:1981}. 
(For various theoretical developments see~\cite{Visser:1993, Visser:1998, LRR, Lake-Como,  diagrams, trapped, no-entropy} and~\cite{Barcelo:2003, Barcelo:2001, Barcelo:2000}. 
For a laboratory implementation using surface waves see~\cite{Weinfurtner:2010, Weinfurtner:2013}.  For a more recent laboratory implementation using BECs see~\cite{Steinhauer:2014}.) 

There is widespread agreement that an acoustic horizon (defined by the normal component of fluid velocity exceeding the local speed of sound) will emit an approximately Planckian spectrum of Hawking phonons; but there is absolutely no requirement that the acoustic horizon be an \emph{event} horizon --- in fact by accelerating or decelerating the fluid flow it is easy to make acoustic horizons appear and disappear at will. 
Any horizon that can completely disappear (without any trace of its prior existence, and without any way of permanently hiding correlations) will qualitatively behave like a leaky furnace with a small hole in it, the original 1850's classical thermodynamic definition of a ``black body''. 
So in these analogue systems, not only is there no reason to believe that there is any ``information puzzle'', but in contrast there is every reason to believe that ordinary unitary evolution and standard physics applies. 

Consequently, even if one could somehow prove that the Hawking photons coming from a specifically general relativistic black hole were uncorrelated, this would merely be a side-effect of the specific details of general relativistic black holes, (as opposed to the generic features of analogue black holes); it would have \emph{nothing} to do with the fundamental physics underlying Hawking radiation itself.
In short:
\begin{itemize}
\item 
Hawking radiation is associated with the apparent/trapping horizon, and couldn't care less about the event horizon (if present).
\item
Unitarity violation (if present) is associated with the event horizon (if present), and couldn't care less about the apparent/trapping horizon. 
\end{itemize}

\section{Discussion} 

 In short, the so-called ``information puzzle'', (often somewhat excessively referred to as the ``information paradox''), is intimately reliant on the assumed existence of an \emph{event} horizon, and much of the force of the information puzzle simply goes away once one uses \emph{apparent} horizons or \emph{trapping} horizons to define what we mean by a black hole~\cite{horizons, dublin, observability}. 
This observation  is closely related to Hawking's recent arguments regarding the necessity of making careful physical distinctions between the mathematical concepts of \emph{event} horizon and \emph{apparent} horizon~\cite{horizons}: 
``The absence of event horizons means that there are no black holes --- in the sense of regimes from which light can't escape to infinity. 
There are, however, apparent horizons which persist for a period of time.'' 
Similarly, a decade ago Hawking asserted~\cite{dublin}: 
``The way the information gets out seems to be that a true event horizon never forms, just an apparent horizon.''

The physical picture that then emerges matches quite nicely with certain proposals for the Hawking radiation process, both somewhat older and more recent, that make no intrinsic reference to \emph{event} horizons per se~\cite{Roman, Parikh:1998, Ashtekar, Hayward05a, Hayward05b, micro-survey, Hossenfelder:2009, Frolov:2014, Israel:2014, Bardeen:2014}.   
Exact thermality of the Hawking flux, and a total absence lack of correlations in the Hawking flux, is often asserted in the scientific literature --- but neither assertion holds up to any level of scrutiny. 
The spectrum of the Hawking flux is certainly not exactly Planckian, and the effective temperature (suitably defined) of the Hawking flux is not equal to, but is instead bounded above, by the Hawking temperature. Whether or not correlations exist in the Hawking flux is contingent upon the assumed existence of event horizons (as opposed to apparent/ trapping horizons). 
Certainly event horizons are not necessary for the development of a Hawking flux, and the often assumed survival of classical event horizons in semi-classical physics is an assumption that is increasingly in doubt~\cite{observability, horizons, dublin}.

There is a crucial difference between the ``qualitative" and ``quantitative" information loss problems.
\enlargethispage{20pt}
\begin{itemize}
\item 
The ``qualitative" problem is this: If a spacelike singularity forms (in the strict mathematical sense), then there will be a (strict mathematical) event horizon,  and unavoidably \emph{some} loss of unitarity associated with any matter that might cross the \emph{event} horizon. 
\item 
The ``quantitative" problem is this: \emph{How much} information is lost behind the event horizon, (if it forms), and how much comes out in the Hawking radiation?\footnote{
Very recent articles specifically addressing this specific point include those by Brustein and Medved~\cite{Medved:2014a, Medved:2014b}, and by Saini and Stojkovic~\cite{Saini:2015}. 
In these articles it is the \emph{quantitative estimates} of the information budget that are important.
} 
\end{itemize}
Only if you \emph{assume} that the event horizon actually forms, \emph{and that it closely tracks the apparent/trapping horizon},
is there ever any significant information loss.


\section*{Acknowledgments}

This research was supported by the Marsden Fund, and by a James Cook fellowship, both administered by the Royal Society of New Zealand.



\end{document}